\documentclass[aps,twocolumn,nofootinbib]{revtex4}
\usepackage{graphicx}
\usepackage{amsmath,amssymb}
\usepackage[colorlinks,citecolor=blue,linkcolor=blue,urlcolor=blue]{hyperref}
\begin{document}
\title{Death and resurrection of the zeroth principle of thermodynamics}
\author{Hal~M.~Haggard}
\author{Carlo Rovelli}
\affiliation{Centre de Physique Th\'eorique de Luminy, Aix-Marseille Universit\'e, F-13288 Marseille, EU}
\date{\small\today}
\begin{abstract}
\noindent The zeroth principle of thermodynamics in the form ``temperature is uniform at equilibrium" is notoriously violated in relativistic gravity. Temperature uniformity is often derived from the maximization of the total number of microstates of two interacting systems under energy exchanges. Here we discuss a generalized version of this derivation, based on informational notions, which remains valid in the general context. The result is based on the observation that the time  taken by any system to move to a distinguishable (nearly orthogonal) quantum state is a universal quantity that depends solely on the temperature. At equilibrium the net information flow between two systems must vanish, and this happens when two systems transit the same number of distinguishable states in the course of their interaction. 
\end{abstract}
\maketitle

\section{Non-uniform equilibrium temperature}

According to non-relativistic thermodynamics, a thermometer (say, a line of mercury in a glass tube), moved up and down  a column of gas at equilibrium in a constant gravitational field, measures a \emph{uniform} temperature.  But this prediction is wrong. Relativistic effects  make the gas warmer at the bottom and cooler at the top, by a correction proportional to $c^{-2}$, where $c$ is the speed of light. This is the well known Tolman-Ehrenfest effect, discovered in the thirties \cite{Tolman:1930zz,TolmanEhrenfest} and later derived in a variety of different manners \cite{Tolman, Landau, Balazs, Ehlers,Tauber,Balazs2,Buchdahl,Ebert,Stachel}. The temperatures $T_{1}$ and $T_{2}$  measured by the same thermometer at two altitudes $h_{1}$ and $h_{2}$ in a Newtonian potential $\Phi(h)$ are related by the Tolman law
\begin{equation}
T_{1}\left(1+\frac{\Phi(h_{1})}{c^{2}}\right)=T_{2}\left(1+\frac{\Phi(h_{2})}{c^{2}}\right).
\label{uno}
\end{equation}
The general-covariant version of this law reads 
\begin{equation}
T|\xi|= constant, 
\label{2}
\end{equation}
where $|\xi|$ is the norm of the timelike Killing field with respect to which equilibrium is established.   

A violation of the uniformity of temperature seems counterintuitive at first, especially if one has in mind a definition of ``temperature" as a label of the equivalence classes of all systems in equilibrium with one another.  In a relativistic context a physical thermometer does not measure this label and we must therefore distinguish two notions: \emph{(i)} a quantity $T_{o}$ defined as this label (proportional to the constant in \eqref{2}), and \emph{(ii)} the temperature $T$ measured by a standard thermometer. 

In  the micro-canonical framework the entropy $S(E)$ is the logarithm of the number of microstates $N(E)$ that have energy $E$ and
 $T$ can be identified with the inverse of the derivative of $S(E)$,
\begin{equation}
\frac{dS(E)}{dE}=\frac1{kT}, 
\end{equation} 
where $k$ is the Boltzmann constant.  The fact that two systems in equilibrium have the same $T$ can be derived by maximizing the total number of states $N=N_{1}N_{2}$ under an energy transfer $dE$ between the two. This gives easily $T_{1}=T_{2}$.  In the presence of relativistic gravity, this derivation fails because conservation of energy becomes tricky: intuitively speaking, the energy $dE$ reaching the upper system is smaller than the one leaving the lower system because ``energy weighs". 

Is there a more general statistical argument that governs equilibrium in a relativistic context? Can the Tolman law be derived from a principle generalizing the maximization of the number of microstates, without recourse to specific models of energy transfer, as is commonly done in the derivations of the Tolman-Ehrenfest effect? 

In this paper we show that the answer to these questions is positive, and we provide a generalization of the statistical derivation of the uniformity of temperature, which remains valid in a relativistic context.   

The core idea is to focus on \emph{histories} rather than \emph{states}.  This is in line with the general idea that states at fixed time are not a convenient handle on general relativistic mechanics, where the notion of \emph{process}, or history, turns out to be more useful  \cite{Rovelli:2004fk}.   Equilibrium in a stationary spacetime, namely the Tolman law, is our short-term objective, but our long-term aim is understanding equilibrium in a fully generally covariant context, where thermal energy can  flow also to gravity \cite{Rovelli:1993ys,Connes:1994hv,Rovelli:2012nv}, therefore we look for a general principle that retains its meaning also in the absence of a background spacetime. 

We show in this paper that one can assign an information content to a history, and two systems are in equilibrium when their interacting histories have the same information content.  In this case the net information flow vanishes, and this is a necessary condition for equilibrium.  This generalized principle reduces  to standard thermodynamics in the non-relativistic setting, but yields the correct relativistic generalization.

This result is based on a key observation: at temperature $T$, a system transits  
\begin{equation}
\tau=\frac{kT}{\hbar}t
\end{equation}
states in a (proper) time $t$, in a sense that is made precise below.  The quantity $\tau$ was introduced in  \cite{Rovelli:1993ys,Connes:1994hv} with different motivations, and called \emph{thermal time}. Here we find the physical interpretation of this quantity: it is time measured in number of elementary ``time steps", where a step is  the characteristic time taken to move to a distinguishable quantum state. Remarkably, this time step is universal at a given temperature.  Our main result is that two systems are in equilibrium if during their interaction they cover the same number of time steps. 

\section{The universal time step}

\begin{figure}
\centerline{\includegraphics[height=5.5cm]{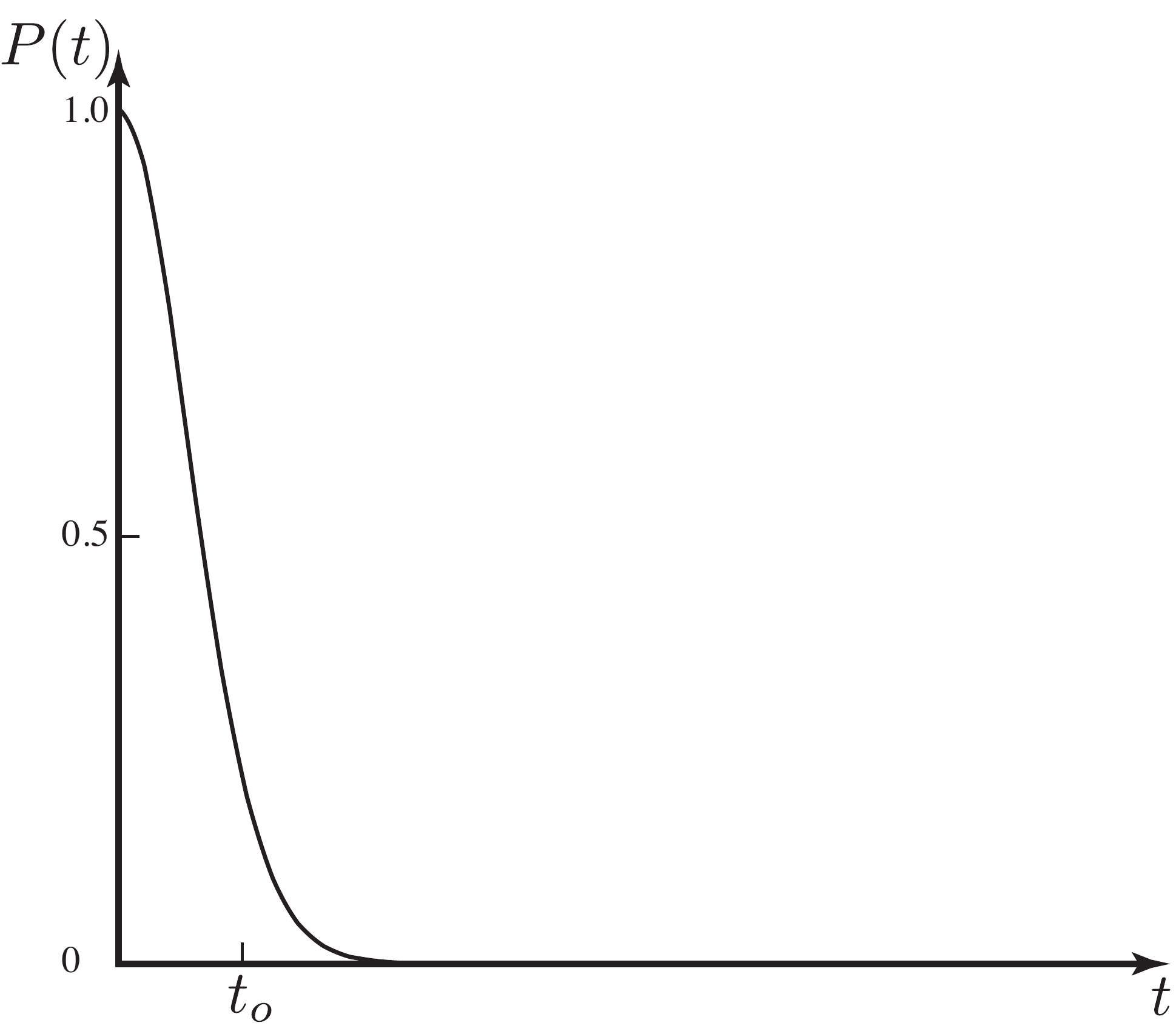}}
\caption{Typical overlap between $\psi(0)$ and $\psi(t)$ as a function of time.}
\label{one}
\end{figure}

Consider a conventional hamiltonian system with hamiltonian operator $H$.  Let $\psi(0)$ be the state at time zero and $\psi(t)$ its evolution. What is the time scale for  $\psi(t)$ to become significantly distinct from  $\psi(0)$? The separation of the state from its initial position is given by the overlap between  $\psi(0)$ and  $\psi(t)$, namely 
\begin{equation}
P(t)=|\langle   \psi(0)|\psi(t)  \rangle |^{2}.
\end{equation}
The typical behavior of $P(t)$, for instance in the case of a semiclassical wave packet, is as in Figure \ref{one}.
The state becomes rapidly distinguishable (nearly orthogonal) to the initial state, in a short time.  Let us call $t_o$ the characteristic decay time for the system self overlap. What is its value? The time  $t_o$ can be estimated by Taylor expanding $P(t)$ for small times.  The first time derivative of $P(t)$ clearly vanishes at $t=0$ which is a maximum, therefore we get the time scale from the second derivative. A straightforward calculation gives
\begin{equation}
\frac{d^{2}P(t)}{dt^{2}}=\frac1{\hbar^{2}}(\langle H^{2}\rangle-\langle H\rangle^{2})=\frac{(\Delta E)^{2}}{\hbar^{2}},
\end{equation}
which implies a characteristic decay time 
\begin{equation} 
t_{o}=\frac{\hbar}{\Delta E},\label{step}
\end{equation}
in accord with the time-energy Heisenberg principle, and with the fact that energy eigenstates ``do not change".\footnote{Notice that this observation provides also a direct meaning to the time-energy uncertainty principle.} 

\begin{figure}
\centerline{\includegraphics[height=3.5cm]{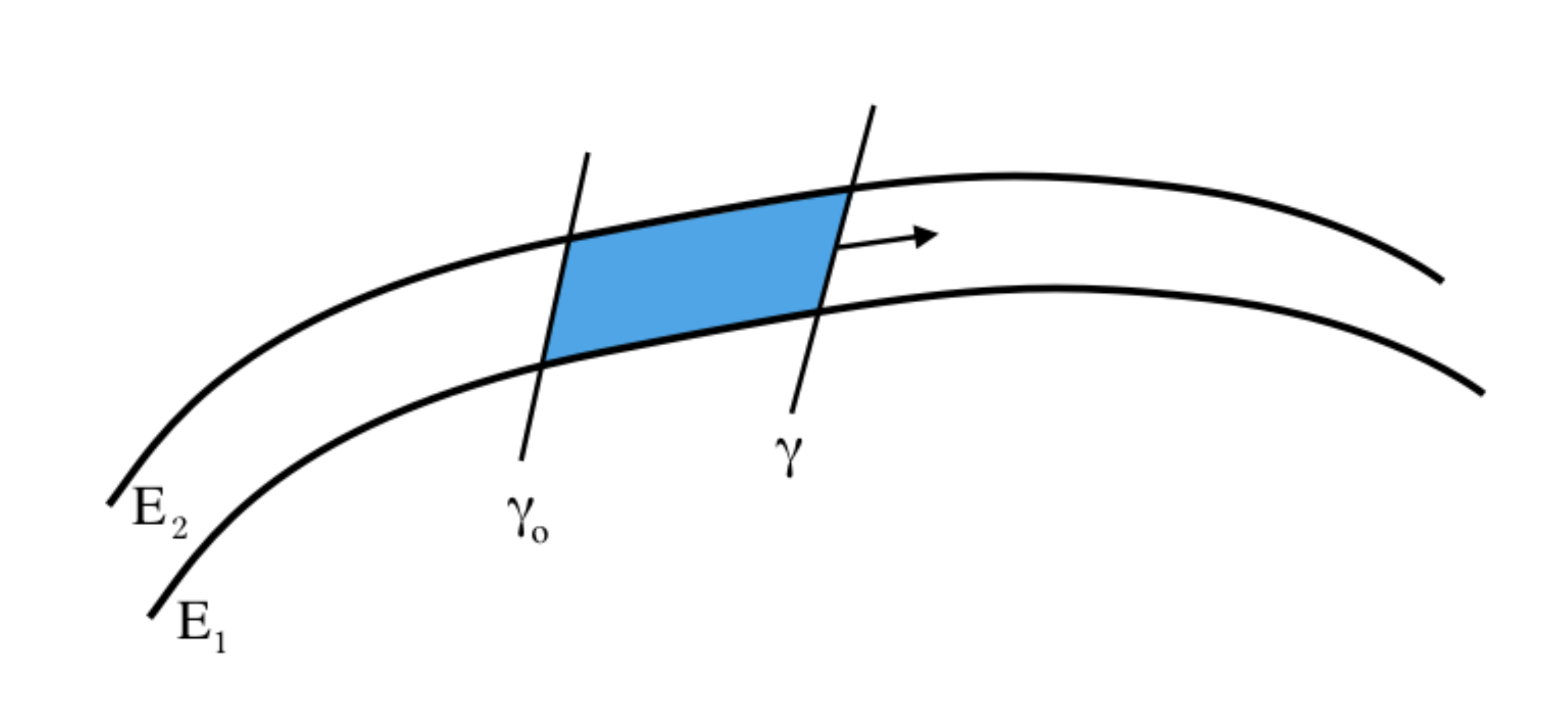}}
\caption{A phase space region moves from one cell to next in the time $\tau_0\sim h/\Delta E$.}
\label{due}
\end{figure}

The same conclusion can be reached also in the classical theory.  Consider a classical hamiltonian system with phase space $\Gamma$ and hamiltonian $H$. For simplicity, say that the system has a single degree of freedom, so that $\Gamma$ is two-dimensional. Let $E_{1}$ and $E_{2}$ be two (nearby) equal-energy surfaces and $\gamma$ a line joining the two surfaces. Consider the motion of $\gamma$ under the time flow, see Figure 2. How long does it take for $\gamma$ to sweep a (small) phase-space volume $V$? The answer is easy to find: the volume of the region $R$ swept by $\gamma$ is the integral of the symplectic two form $\omega=dp\wedge dq$ and its time derivative is
\begin{equation}
\frac{dV(t)}{dt}=\frac{d}{dt}\int_{R(t)}\omega=\int_{\gamma}\omega(X),
\end{equation}
where $X$ is the hamiltonian time flow. This is given by the Hamilton equations, which can be written in the compact form
\begin{equation}
\omega(X)=-dH.
\end{equation}
Inserting this in the previous equation gives
\begin{equation}
\frac{dV(t)}{dt}=-\int_{\gamma}dH=E_{2}-E_{1}\equiv\Delta E.
\end{equation}
Now consider a small region of phase space, such as the blue region in Figure \ref{due}. Say that the volume of this region is $\hbar$. How long does it take for this region to be carried along by the dynamics to a new position where the overlap with its initial location is negligible?  It is clear that the answer is again \eqref{step}.

This is the same result as in the quantum theory: the time step $t_{o}$ is the time taken generically to move from a state to a distinct (orthogonal) state.  
Indeed, a semiclassical state can be viewed as related to a Planck-size cell of the classical phase space, and  the decay time of the quantum overlap $P(t)$ is essentially the time the system moves from one  Planck-size cell to the next.  The argument can be repeated with a bit more labour for a system with many degrees of freedom.\footnote{If  we can diagonalize locally the state and the dynamics in energy-angle variables $(E_n,\phi_n)$, the phase space volume swept by the boundary of a given region, in a time $dt$, is
$
dV \sim \sum_n \Delta E_{n}({V}/{V_n}) dt, 
$
where $V_n$ is the phase space volume of the $n$-th degree of freedom. A coherent state has volume $\hbar$ in each of its degrees of freedom, giving 
$
dV \sim \hbar^{n-1} \sum_n {\Delta E_{n}} dt \sim  \hbar^{n-1}\Delta E dt. 
$ Since a phase space cell has volume $\hbar^n$, the time taken to move one cell is again $\sim \hbar/\Delta E$.
}

Let us now consider a system in thermal equilibrium with a thermal bath at temperature $T$.  Its mean energy is going to be $kT$ and the variance of the energy is also going to be $kT$. Thus we have $\Delta E \sim kT$. At a given temperature $T$, consider the time step
\begin{equation}
t_{o}=\frac{\hbar}{kT}.
\label{to}
\end{equation}
According to the previous discussion, this  is the average time the system takes to move from a state to the next (distinguishable) state. This average time step is therefore universal: it depends only on the temperature, and not on the properties of the system.

\section{Thermal time, temperature and their physical meaning}

The dimensionless quantity 
\begin{equation}
\tau=\frac{t}{t_{o}} 
\end{equation}
measures time in units of the time step $t_{o}$, that is, it estimates the number of distinguishable states the system has transited during a given interval. For a system in thermal equilibrium, \eqref{to} gives
\begin{equation}
\tau=\frac{kT}{\hbar} t. 
\label{tau}
\end{equation}
This same quantity was introduced with  different motivations in \cite{Rovelli:1993ys,Connes:1994hv} under the name \emph{thermal time}. It is the parameter of the Tomita flow on the observable algebra, generated by the thermal state.  In the classical theory, it is the parameter of the hamiltonian flow of $h=-\ln \rho$, where $\rho$ is a Gibbs state, in $\hbar=k=1$ units.   

The argument in the previous section unveils the physical interpretation of thermal time: thermal time, which is dimensionless, is  simply the number of distinguishable states a system has transited during an interval. In a sense, it is  ``time counted in natural elementary steps", which exist because the Heisenberg principle implies an effective granularity of the phase space. 

Notice also that temperature is the ratio between  thermal time and (proper) time \cite{Rovelli:2010mv}
\begin{equation}
T =  \frac{\hbar\tau}{k  t} . 
\end{equation}
Accordingly, in  $\hbar=k=1$ units temperature is measured in ``states per second" and is nothing other than the number of states transited by the system per unit of (proper) time. This is the general informational meaning of temperature. A warmer system is a system where individual states move faster across unit cells of phase space. 

\section{Equilibrium between histories}

Let us come to the notion of equilibrium. Consider two systems, System 1 and System 2, that are in interaction during a certain interval. This interaction can be quite general but should allow the exchange of energy between the two systems. During the interaction interval the first system transits $N_{1}$ states, and the second $N_{2}$, in the sense illustrated above. Since an interaction channel is open, each system has access to the information about the states the other has transited through the physical exchanges of the interaction. 

The notion of \emph{information} used here is purely physical, with no relation to semantics, meaning, significance, consciousness, records, storage, or mental, cognitive, idealistic or subjectivistic ideas. Information is simply a measure of a \emph{number of states}, as is defined in the classic text by Shannon \cite{Shannon:1948fk}. 

System 2 has access to an amount of information $I_1=\log N_{1}$ about System 1, and System 1 has access to an amount of information $I_2=\log N_{2}$ about System 2. Let us define the net flow of information in the course of the interaction as $\delta I=I_2-I_1$.  Equilibrium is by definition invariant under time reversal, and therefore any flow must vanish. It is therefore interesting to \emph{postulate} that also the information flow $\delta I$ vanishes at equilibrium.  Let us do so, and study the consequences of this assumption. That is, we consider the possibility of taking the vanishing of the information flow 
\begin{equation}
\delta I = 0 
\label{equi}
\end{equation}
as a general condition for equilibrium, generalizing the maximization of the number of microstates of the non-relativistic formalism.\footnote{In the micro-canonical framework  equilibrium is characterized by maximizing entropy, namely the number of micro-states sharing given macroscopic values. This is meaningful, e.g. under the ergodic hypothesis, according to which time averages can be replaced by phase-space averages. In other words, if  the ergodic hypothesis holds, a micro-canonical ensemble is essentially the family of states over which the single real individual microstate wanders. What we are doing here is essentially undoing this step and moving back from phase-space ensembles to actual histories. In the classical theory, there is a measure associated to a spacetime volume and not to the length of a history. But in this paper we have shown that there is also a natural measure associated to the history of a quantum state. This allows us to backtrack from phase-space volume to number of steps along the history.}

Let us see what this implies. At equilibrium
\begin{equation}
N_{1}=N_{2}.
\label{fp}
\end{equation}
Since the rate that states are transited is given by $\tau$ and we assume a fixed interaction interval, the equilibrium conditions also reads
\begin{equation}
\tau_{1}=\tau_{2}. \label{tau1tau2}
\end{equation}
Now, consider a non-relativistic context where two system are  in equilibrium states at temperatures $T_{1}$ and $T_{2}$, respectively. In the non-relativistic limit, time is a universal quantity, which we call $t$.  Therefore the condition \eqref{tau1tau2} together with \eqref{tau} implies that $t_{o} = \hbar/kT$ has the same value for the two systems and  $T_{1}=T_{2}$, which is the standard non-relativistic condition for equilibrium: temperature is uniform at equilibrium.   On a curved spacetime, on the other hand, (proper) time is a local quantity $ds$ that varies from one spatial region to another. Therefore thermal time is given  by
\begin{equation}
d\tau=\frac{kT}{\hbar} ds. 
\end{equation}
In order for equilibrium to exist on a given spacetime, spacetime itself must be stationary, namely have a timelike Killing field $\xi$, and an equilibrium configuration will be $\xi$ invariant. Proper time along the orbits of $\xi$ is $ds=|\xi|dt$ where $t$ is an affine parameter for $\xi$. Therefore thermal time is now
\begin{equation}
d\tau=\frac{kT}{\hbar} |\xi| dt.
\end{equation}
If two systems located in regions with different $|\xi|$ are in thermal contact for a finite interval $\Delta t$, then they are in equilibrium if $|\xi|T$ has the same value.  This is precisely the Tolman law \eqref{2}.  Therefore the generalized first principle \eqref{equi} gives equality of temperature in the non relativistic case and the Tolman law in the general case.

In static coordinates, $ds^{2}=g_{00}(\vec x)dt^{2}-g_{ij}(\vec x)x^{i}x^{j}$ and thermal time is proportional to coordinate time. The Killing vector field is $\xi=\partial/\partial t$ and $|\xi|=\sqrt{g_{00}}$. In the Newtonian limit $g_{00}=1+2\Phi/c^{2}$ and we recover \eqref{uno}. 

Returning to the cylinder of gas in a constant gravitational field we see that during a coordinate-time interval $\Delta t$ the proper times lapsed in the upper and lower systems are different:  identical clocks at different altitudes run at different rates. But the lower system is hotter, its degrees of freedom move faster in clock time from one state to the next. This faster motion compensates exactly the slowing down of proper time, so that {\em  upper and lower systems transit the same number of  states during a common interaction interval $\Delta t$.}   While a pendulum slows down in a deeper gravitational potential, at equilibrium all systems transit from state to state at the same common rate, independent from the gravitational potential.  This result provides a simple and intuitive interpretation of the Tolman effect.

\section{Wien's displacement law}
The Tolman-Ehrenfest effect is a small relativistic correction, at the surface of the earth $\nabla T/T = 10^{-18} \text{cm}^{-1}$, and is not yet experimentally established. The principle proposed here also provides a mellifluous derivation of the well-tested Wien displacement law. 

Consider an isothermal cavity filled with electromagnetic radiation. A slow, adiabatic expansion of the cavity leaves the radiation in equilibrium throughout the expansion process. During this expansion both the normal mode frequencies and the temperature of the radiation are adjusted. For the mode of frequency $\nu$, the condition of remaining in equilibrium is that the slow expansion take place on a time scale much greater than the period of the mode, i.e. $t_{\text{exp}}>> t_{\nu} \equiv 1/\nu$. Hence, the relevant clock for this mode is its period. 

The condition that this mode be in equilibrium during the entire expansion \textit{history} is that 
\begin{equation}
\tau = \frac{kT}{\hbar} t_{\nu}= \text{const.}
\end{equation}
or expressed in terms of the mode's frequency,
\begin{equation}
\frac{T}{\nu} = \text{const.}, 
\end{equation}
which is precisely the general form of Wien's displacement law. This special relativistic application of the principal proposed here plays an important role in the astrophysical determination of star temperatures.

\section{Conclusions}

We have suggested a generalized statistical principle for equilibrium in statistical mechanics.   We expect that this will be of use going towards a genuine foundation for general covariant statistical mechanics.  

The principle is formulated in terms of histories rather than states and expressed in terms of information. It reads: \emph{Two histories are in equilibrium if the net information flow between them vanishes, namely if they transit the same number of states during the interaction period}. 

This is equivalent to saying that the thermal time $\tau$ elapsed for the two systems is the same, since thermal time is the number of states transited, or, equivalently, is (proper) time in $t_{o}$ units, where $t_{o}$ is the (proper) time needed for a system to transit to an orthogonal state. 
The elementary (proper) time step $t_{o}$ is given by $\hbar/kT$ and is a universal quantity for all systems at temperature $T$.

In non-relativistic physics, time is universal and the above principle implies that temperature is uniform at equilibrium. On a curved spacetime, proper time varies locally and what is constant is the product of temperature and proper time. 

Temperature admits the informational interpretation as states transited per second, consistent with the fact that in $\hbar=k=1$ units it has dimension of $second^{-1}$.  Temperature is the rate at which systems move from state to state.

\centerline{------}

\vspace{.5em}
Thanks to Badr Albanna and Matteo Smerlak for carefully reading the manuscript and to Alejandro Perez and Simone Speziale for useful discussions. H. M. H. acknowledges support from the National Science Foundation (NSF) International Research Fellowship Program (IRFP) under Grant No. OISE-1159218.


\end{document}